\documentclass[reprint,amsmath,amssymb,aps,pra]{revtex4-1}

\usepackage{graphicx}
\usepackage{bm}
\usepackage{amsmath}
\usepackage{amsfonts}
\usepackage{amssymb}
\usepackage[latin1]{inputenc}

\newcommand{\mbf}[1]{\mathbf{#1}}

\newcommand{\ket}[1]{|#1\rangle}
\newcommand{\bra}[1]{\langle#1|}

\newcommand{\scalar}[2]{\langle#1|#2\rangle}

\newcommand{\op}[1]{|#1\rangle\langle#1|}
\newcommand{\opp}[2]{|#1\rangle\langle#2|}

\newcommand{\bvec}[1]{\mathbf{#1}}

\begin{document}
\title{Asymptotic reduced density matrix of discrete-time quantum walks}
\author{Mostafa Annabestani}
\email{Annabestani@shahroodut.ac.ir}
\affiliation{Faculty of Physics, Shahrood university of technology, Shahrood, Iran}
\date{\today}
\begin{abstract}
In this article we show that for any quantum walker with \textit{m}-dimensional coin subspace, we have $m^2\times m^2$ specific constant matrix $\mathcal{C}$ where it completely determines the asymptotic reduced density matrix of the walker. We show that for any initial state with $P_0$ projector, reduced density matrix, can be obtained by $Tr_1\left(P_0\otimes I\;\mathcal{C}\right)$ or equivalently $Tr_2\left(I\otimes P_0\;\mathcal{C}\right)$. It is worth to mention that characteristic matrix $\mathcal{C}$  is independent of the initial state and just depends on coin operator, so by finding this matrix for specific type of QW the long-time behavior of it, such as local state of the coin after a long time walking and asymptotic entanglement between coin and position will be completely known for any initial state. We have found the characteristic matrix $\mathcal{C}$ for general coin operator, $U\left(2\right)$, as well as exact form of this matrix for local initial state.
\end{abstract}


\maketitle

\section{Introduction}
\label{sec:intro}
Quantum walks (QW)\cite{Aharonov-1993}  are the quantum counterparts of classical random walk. While the degrees of freedom of a random variable determines the direction of the walk in classical random walk, in quantum walk the state of a quantum system determines the direction of the walk. Since the quantum system can be in superposition state, quantum walker can walk in different directions simultaneously. 

The existence of quantum features which do not have a classical counterpart, such as superposition, quantum interference and entanglement, has made quantum walks to have different properties and in many cases to be much more powerful. These differences and potentials in quantum walk have made it an interesting topic for research.
Some studies have illustrated the power of QW as the universal computation approach \cite{Childs2009,Lovett-2010} while others have introduced quantum walk based quantum algorithms \cite{Childs04,Shenvi,Feldman2010}. The power and applications of QW motivated researchers to introduce different models of QWs. Many coins QW \cite{Brun2003ManyCoin}, QW on higher dimension \cite{Mackay}, QW on a circle \cite{cycle1}, coin less QW \cite{Patel2005}, Mobius QW \cite{Moradi2017} and staggered QW \cite{Portugal2016} are just a few examples of these models.

Furthermore, some important phenomena such as the localization \cite{Inui2004Localization,Vakulchyk2017localization}, topological phases \cite{Kitagawa2010,Wang2018TopologicalPhase} and decoherency \cite{Brun1,Annabestani2010} have been studied well in/by QWs.

The study of asymptotic behavior of QW is another field that researchers are interested in. Asymptotic coin-position entanglement (CPE) \cite{annabestani,Orthey2017}, global chirality distribution (GCD) \cite{Romanelli2010Charality} or Thermodynamic behavior of QW \cite{Romanelli2012Thermodynamic} are a few examples in which asymptotic behavior of QW have been studied. Clearly to study asymptotic behavior, the numerical methods are not suitable or at least are not accurate, so we need exact analytical method. 
Although many analytical papers have been published, they are usually restricted to specific situations (specific initial state or coin operator) \cite{Romanelli2010Charality,Diaz2016,Romanelli2012Thermodynamic,Abal2006,Orthey2017}), because analytical calculations in general form are very tedious and difficult.

In this paper, we introduce a general method for calculating a reduced density matrix of QW that not only simplifies the calculation but it is also applicable to the general case. In this method, for each $m$-dimensional quantum walk, we obtain a $m^2\times m^2$ characteristic matrix (we call it reduced density characteristic matrix (RDCM)) that contains all the information about the density matrix  of QW and reduced density matrix  will be obtained by a simple formula for any desired initial state. 

Furthermore we calculate RDCM for most general form of 1DQW in this paper as well. Henceforth, the reduced density matrix of 1DQW will be fully determined for the most general coin operator and any initial state.

It should be noted that the RDCM for 1DQW is just an example of our formalism and the reader can obtain the RDCM for other models of QWs. The good news is that RDCM  is calculated once for each model, and then the density matrix can be easily calculated for any initial state in that model.

This paper is organized as follows: in section II, by introducing $m$-dimensional quantum walk, we have derived general form of asymptotic reduced density matrix of $m$-dimensional based on RDCM. In section III we have shown that for two family of initial states, our formalism became more simplified. In section IV, we have calculated RDCM for most general form of coin operator in 1DQW and brought some examples in section V.The conclusion have presented in the last section.

\section{formalism}
\label{sec:formalism}
The \textit{m}-dimensional quantum walk can be defined in terms of discrete \textit{m}-dimensional space whose sites are labeled by vector $\mbf{r}=(r_0,...,r_{m-1})$. The set of orthonormal states $\{\ket{\mbf{r}}\}$ span the position subspace, ${\cal H}_P$, of the walker. The ``coin'' degree of freedom is represented by a \textit{m}-qubit space, ${\cal H}_C$, spanned by $2^m$ orthonormal states which we label as $\left\lbrace\ket{j}\mid j=j_{m-1}...j_1j_0 \text{ is bitstring}\right\rbrace$. This nomenclature is motivated by the quantum walk on a line, where $\ket{L}=\ket{0}=(1,0)^T$ and $\ket{R}=\ket{1}=(0,1)^T$ are associated with left or right displacements respectively. In our generalization, we use the bit $j_i$ to define of  movement's direction along axis $r_i$.

The Hilbert space for the system is ${\cal H}={\cal H}_P\otimes{\cal H}_C$ and a generic state is
\begin{equation}\label{eq:gen_state}
 \ket{\Psi}=\sum_{\mbf{r}}\sum_{j=0}^{2^m-1} a_j(\mbf{r})\, \ket{\mbf{r}}\otimes\ket{j}.
\end{equation}
One step of evolution is described by
\begin{equation}\label{eq:evol_generic}
 \ket{\Psi(t+1)}=U\ket{\Psi(t)},
\end{equation}
where
\begin{equation}\label{eq:Uop}
 U=S (I_P\otimes U_C),
\end{equation} 
in which $I_P$ is the identity operator in ${\cal H}_P$. The evolution combines a unitary coin operation $U_C$ in ${\cal H}_C$ with a shift operator
\begin{equation}\label{general_S}
  S = \sum_{\mbf{r}}\sum_{j}\opp{\mbf{r}+\mbf{s}_j}{\mbf{r}}\otimes\op{j}
\end{equation}
where $\mbf{s}_j$ is usually \textit{m}-dimensional normal vector ($\left|\mbf{s}_j\right|$ defines the length of steps. So non-normal $\mbf{s}_j$ defines QW with non-unit length of steps). \textit{S} performs the conditional displacements determined by the coin state \textit{j}. The correspondence between coin states and displacements is not unique. The different definition of $\mbf{s}_j$ will define different family of QWs. 

From \eqref{eq:evol_generic} it is clear that, the state of system after $t$ steps is
\begin{equation}\label{evol_t}
\ket{\Psi(t)}=U^t\ket{\Psi(0)},
\end{equation} 
in which $\ket{\Psi(0)}$ is the initial state. Unfortunately calculation of $U^t$ is not easy, because $U$ is not diagonal in basis $\lbrace \ket{\mbf{r}} \rbrace$ (see \eqref{eq:Uop} and \eqref{general_S})


The Fourier transform, as first noted in this context by Nayak and Vishwanath \cite{Nayak}, is extremely useful because shifting operator $S$ is diagonal in $k$-space. By Fourier transformation we can change the basis to

\begin{align}\label{FourierTransform}
 \ket{\mbf{k}}&=\sum_r  e^{i\mathbf{k}\cdot \mbf{r}}\ket{\mbf{r}}
\end{align}
where $\mathbf{k}$ is a vector with real continuous components $(k_0,...,k_{m-1})$ in the interval $[-\pi,\pi]$. We are always able to back to the original basis $\lbrace \ket{\mbf{r}}\rbrace$ by inverse Fourier transformation
\begin{align}\label{InvFourierTransform}
\ket{\mbf{r}}&=\int{\frac{d^m\mbf{k}}{\left(2\pi\right)^m} e^{-i\mathbf{k}\cdot \mbf{r}}\ket{\mathbf{k}}}.
\end{align}

From \eqref{eq:evol_generic}, each component of wave vector, in $k$-space evolves as
\begin{align}\label{k_waveVector}
\scalar{\mbf{k}}{\Psi(t+1)}=\int{\frac{d^m\mbf{k'}}{\left(2\pi\right)^m} \bra{\mathbf{k}}U\ket{\mathbf{k^\prime}}\scalar{\mathbf{k'}}{\Psi(t)}}.
\end{align}
in which we use the completeness of basis $\lbrace \ket{\mbf{k}}\rbrace$,
\begin{align}
\int{\frac{d^m\mbf{k'}}{\left(2\pi\right)^m} \ket{\mathbf{k'}}\bra{\mathbf{k'}}}=I.
\end{align} 
By using \eqref{eq:Uop},\eqref{general_S} and \eqref{InvFourierTransform}
\begin{align}
\bra{\mathbf{k}}U\ket{\mathbf{k^\prime}}=\sum_{\mbf{r}}\sum_{j}{e^{-i\left(\mbf{k}-\mbf{k'}\right)\cdot \mbf{r}}}{e^{-i\mbf{k}\cdot\mbf{s}_j}}\op{j}U_C.
\end{align}
and using the definition of
\begin{align}
\delta\left(\mbf{k}-\mbf{k'}\right)=\frac{1}{\left(2\pi\right)^m}\sum_{\mbf{r}}{e^{-i\left(\mbf{k}-\mbf{k'}\right)\cdot \mbf{r}}},
\end{align}
we can write the simple form of \eqref{k_waveVector} as
\begin{equation}\label{1step-k}
\ket{\psi_k(t+1)}=U_k\ket{\psi_k(t)}
\end{equation}
in which
\begin{equation}\label{general_U_k}
	U_k=\sum_j{e^{-i\mbf{k}\cdot\mbf{s}_j}\opp{j}{j}}U_C.
\end{equation}
and
\begin{equation}\label{psi_k}
\ket{\psi_k(t)}\equiv\scalar{\mbf{k}}{\Psi(t)}.
\end{equation}

This equation defines a unitary operator $U_k$ which is represented by a $2^m\times 2^m$ matrix. The basic idea behind our approach is to use the spectral decomposition of $U_k$ to obtain information about the long-time evolution of the system. 

Let us consider the eigenproblem for the unitary operator $U_k$ with eigenvalues $e^{i\omega_k}$ and corresponding normalized eigenvectors $\ket{\omega_k}$. So by using the spectral decomposition of $U_k$, the state of the system after $t$ steps is
\begin{equation}\label{k-evol}
\ket{\psi_k(t)}= U_k^t\ket{\psi_k(0)}=\sum_{\{\omega_k \}}{e^{i\omega_k t} \ket{\omega_k}\scalar{\omega_k}{\psi_k(0)}},
\end{equation}
where the sum is over the set of eigenvalues of $U_k$.

The aim of this article is to find analytic compact formula for reduced density matrix of the walker after a long time walking. The reduced density matrix is in fact the local state of coin subspace which appears in some parts of quantum information (QI) and quantum computation (QC) processes. Thermodynamical behavior of QW \cite{Romanelli2012,Diaz2016}, Chirality behavior and non-Markovianity in QW \cite{Romanelli2010,Romanelli2014} and coin-position entanglement (CPE) \cite{annabestani} are just some examples of researches which deal with the asymptotic behavior of QW by investigating the reduced density matrix.

From \eqref{evol_t}, the reduced density matrix $\rho_c$ after $t$ steps is defined by 
\begin{equation}
\rho_c(t)=Tr_x\left(\rho(t)\right)=\sum_x{\scalar{x}{\Psi(t)}\scalar{\Psi(t)}{x}}. 
\end{equation}
Although, by using the inverse Fourier transformation,  we are always able to calculate  $\scalar{x}{\Psi(t)}$ from \eqref{k-evol}, but we can avoid this issue by using Parseval's theorem. So
\begin{equation}\label{eq:rho_c1}
\rho_c(t)=\sum_x{\scalar{x}{\Psi(t)}\scalar{\Psi(t)}{x}}=\int\frac{d^m\mbf{k}}{\left(2\pi\right)^m}\, \op{\psi_k(t)},
\end{equation}
where $\mbf{k}\equiv(k_0,k_1,...,k_{m-1})$ with $k_i \in \left[ { - \pi ,\pi } \right]$ and $\protect{d^m\mathbf{k}\equiv dk_0\,dk_1...dk_{m-1}}$.

By putting \eqref{k-evol} in \eqref{eq:rho_c1},
\begin{eqnarray}\label{op-psik}
\rho_c(t)=\int\frac{d^m\mbf{k}}{\left(2\pi\right)^m}\sum_{\{\omega_k,\,\omega_k^\prime\}} e^{i(\omega_k - \omega_k^\prime)t}\; \opp{\omega_k}{\omega_k^\prime} \times\\\nonumber
\scalar{\omega_k}{\psi_k(0)}\scalar{\psi_k(0)}{\omega_k^\prime}.
\end{eqnarray}
In the asymptotic limit $t\gg 1$, according to the stationary  phase theorem, only terms with $\omega_k=\omega_k^\prime$ contribute to eq.~(\ref{eq:rho_c1}) as discussed in detail in Ref.~\cite{Nayak}. Thus
\begin{align}\label{eq:rho_c}
 \hat\rho_c=\int\frac{d^m\mathbf{k}}{(2\pi)^m} \sum_{\{\omega_k\}} \op{\omega_k}\times
\scalar{\psi_k(0)}{\omega_k}\scalar{\omega_k}{\psi_k(0)}
\end{align}
where we use a caret $(\hat~)$ to indicate that the asymptotic limit $\hat\rho_c\equiv\lim_{t\rightarrow\infty} \rho_c(t)$ has been taken.
We can use the linearity of ``trace'' and rewrite Eq.\ref{eq:rho_c} as  
\begin{equation}\label{rho_c-trace-Form}
	\hat\rho_c=\int\frac{d^m\mathbf{k}}{(2\pi)^m} {Tr_1\left(P_0(\mathbf{k})\otimes I\;\mathcal{C}(\mathbf{k})\right)}
\end{equation}
where 
\begin{equation}\label{C}
	\mathcal{C}(\mathbf{k})=\sum_{\{\omega_k\}} \op{\omega_k}\otimes \op{\omega_k},
\end{equation}
$P_0(k)=\op{\psi_k(0)}$, is projector of initial state, and 
$I$ is $m\times m$ identity matrix. Obviously $\mathcal{C}(k)$ is symmetric under the interchanging  part 1 and 2, so $Tr_1\left(P_0(k)\otimes  I\;\mathcal{C}(k)\right)$ can replaced by $Tr_2\left(I\otimes  P_0(k)\;\mathcal{C}(k)\right)$ in \eqref{rho_c-trace-Form}.

As you see, $\mathcal{C}$ is specific summation of cross product of two $m\times m$ operators, therefore $\mathcal{C}$ is $m^2 \times m^2$ matrix which can be considered as a $m\times m$ matrix of block matrix $m\times m$. We use partial trace $Tr_1\left(.\right)$ when we use projector of initial state as $P_0\otimes I$ and it equals to summation of all diagonal blocks of $\mathcal{C}$, similarly  $Tr_2\left(.\right)$ will be applied when we use $I\otimes P_0$ in \eqref{rho_c-trace-Form} in this case each element of result matrix is the trace of corresponding block. Therefore the action of $Tr_1\left(.\right)$ or $Tr_2\left(.\right)$ in the form of \eqref{rho_c-trace-Form} converts the  $m^2 \times m^2$ matrix of $\mathcal{C}$ to $m \times m$ matrix which is reduced density matrix of QW with initial state $\ket{\psi_k(0)}$ after a long time walking.

\section{Initial states}
The \eqref{rho_c-trace-Form} is general and true for any kind of initial states. But for some category of initial states, the problem can be more simplified.
\subsection{Local initial states}
For initial coin state $\ket{\chi}$ which is localized in position $\mathbf{v}$, we have
\begin{equation}
 \ket{\Psi(0)}=\ket{\mathbf{v}}\otimes\ket{\chi_0}.
\end{equation}
So from \eqref{psi_k}
\begin{equation}
\ket{\psi_k(0)}=\scalar{\mbf{k}}{\Psi(0)}= e^{i\mathbf{k}\cdot \mbf{v}}\ket{\chi_0}.
\end{equation}
Therefore
\begin{equation}\label{P0}
P_k(0)=\op{\psi_k(0)}=\op{\chi_0}.
\end{equation}
does not depend on $k$. So from \eqref{rho_c-trace-Form} we can write
\begin{equation}\label{rho_c-local}
\hat\rho_c={Tr_1\left(\op{\chi_0} \otimes I\;\mathcal{C}_L\right)}
\end{equation}
in which
\begin{equation}\label{C_L}
\mathcal{C}_L=\int\frac{d^m\mathbf{k}}{(2\pi)^m} {\mathcal{C}\left(\mathbf{k}\right)}
\end{equation}
is a constant matrix. It should be emphasized that the calculation of $\mathcal{C}\left(\mathbf{k}\right)$ for each specific coin will be performed only once.
 
We will calculate explicit form of ${\mathcal{C}\left(\mathbf{k}\right)}$ and $\mathcal{C}_L$ in section.\ref{formOfC} for the most general form of coin operator, so analytic form of $\hat{\rho_c}$ for any initial state and any coin operator, will be easy to calculate (see \eqref{rho_c-local}). 

\subsection{Non-local separable coin-position initial states}
If we have an initial coin state $\ket{\chi_0}$ distributed over different positions, the coin state can be separated from the position state, therefore
\begin{equation}
\ket{\Psi(0)}=\sum_\bvec{n}{a_\bvec{n}\ket{\bvec{n}}}\otimes\ket{\chi_0}.
\end{equation}
in which $\sum_\bvec{n}{\left|a_\bvec{n}\right|^2}=1$. Therefore
\begin{equation}\label{Qk}
\ket{\psi_k(0)}= \sum_\bvec{n}{a_\bvec{n} e^{i\mathbf{k}\cdot \bvec{n}}\ket{\chi_0}}=Q(\bvec{k})\ket{\chi_0}.
\end{equation}
and
\begin{equation}
P_k(0)=\op{\psi_k(0)}=\left|Q\left(\bvec{k}\right)\right|^2\op{\chi_0}.
\end{equation}
So from \eqref{rho_c-trace-Form} $\hat\rho_c$  can be written as
\begin{equation}\label{rho_c-separable}
\hat\rho_c={Tr_1\left(\op{\chi_0} \otimes I\;\mathcal{C}_S\right)}
\end{equation}
in which
\begin{equation}\label{C_S}
\mathcal{C}_S=\int\frac{d^m\mathbf{k}}{(2\pi)^m} {\left|Q\left(\bvec{k}\right)\right|^2\mathcal{C}\left(\mathbf{k}\right)}.
\end{equation}

 $\mathcal{C}_S$ is a constant matrix, same as $\mathcal{C}_L$, but unlike the local initial state, $\mathcal{C}_S$ depends on $\left|Q\left(\bvec{k}\right)\right|^2$ (distribution on the position space).

\section{General QW on the line with U(2) coin operator}\label{formOfC}
In this section we used the formalism of previous section to find characteristic matrix of $\mathcal{C}(\bvec{k})$ for QW on the line with general form of coin operator $U(2)$.
Although the most general form of $U(2)$ is defined by four parameters $\theta,\,\alpha ,\beta $ and $\phi$ as 
\begin{equation}\label{U(2)definition}
	U\left(2\right)\equiv{\rm e}^{i\frac{\phi}{2}}\left( \begin {array}{cc} {{\rm e}^{i\alpha}}\cos{\theta} &{{\rm e}^{i\beta}}\sin{\theta} 
\\\noalign{\medskip}{-{\rm e}^{-i\beta}}\sin{\theta} &{
{\rm e}^{-i\alpha}}\cos{\theta} \end {array} \right),
\end{equation}
but without loss of generality we can omit $\phi$, because this parameter just multiples eigenvectors by a global phase ${\rm e}^{i\phi/2}$, which is not important in our problem.

Therefore the most general form of the coin operator can be   
\begin{equation}\label{General-UC}
U_C=U\left(\theta,\alpha,\beta\right)=\left( \begin {array}{cc} {{\rm e}^{i\alpha}}\cos{\theta} &{{\rm e}^{i\beta}}\sin{\theta} 
\\\noalign{\medskip}{-{\rm e}^{-i\beta}}\sin{\theta} &{
	{\rm e}^{-i\alpha}}\cos{\theta} \end {array} \right),
\end{equation}
which includes all kinds of coin operators for QW on the line. For example $U\left(\pi/4,\pi/2,\pi/2\right)=H$ (except global phase $-i$) is Hadamard walk and $U\left(\theta,\alpha,\beta\right)$ for $\theta \ne \pi/4$ is unbiased QW and etc.

In one dimensional QW, $\mbf{r}=(r_0)=x$, and two dimensional coin space which determines left and right movements can be defined by $\mbf{s}_1=-1$  and $\mbf{s}_0=1$ respectively. So shifting operator \eqref{general_S} for 1DQW is
\begin{equation}
	S=\sum_x{\opp{x+1}{x}\otimes\opp{0}{0}+\opp{x-1}{x}\otimes\opp{1}{1}}
\end{equation}

According to \eqref{general_U_k} we can write
\begin{equation}
	U_{k}=\left( \begin {array}{cc} {{\rm e}^{-ik}} &0 
\\\noalign{\medskip} 0 &  {{\rm e}^{ik}} \end {array} \right)U\left(\theta,\alpha,\beta\right).
\end{equation} 
The eigenvalues of this unitary matrix is in the form of $e^{i\omega_{k}}$ with $\omega_{k}=\lbrace -\gamma,\gamma \mid \cos{\gamma}=\cos{\theta}\cos{\left(k-\alpha\right)}\rbrace$ and corresponding eigenvectors are
\begin{equation}\label{e_k}
	\ket{\omega_k}=\frac{1}{N}\left( \begin {array}{c} {{\rm e}^{-i\omega_{k}}}-\cos \theta \, {{\rm e}^{-i \left( k
			-\alpha \right) }}
	 \\ \sin \theta\, {{\rm e}^{i \left( k-\beta \right) }}
  \end {array} \right)
\end{equation}\\
in which N is normalization factor. Although, \eqref{e_k} is only thing we need to calculate $\mathcal{C}(\bvec{k})$ from \eqref{C}, but simple substitution \eqref{e_k} into \eqref{C} yields very complex, long and unusable expressions. Fortunately, by several levels of simplifications, we are always able to simplify $\mathcal{C}(\bvec{k})$ to have clean expressions. The success of simplification process directly depends on the complexity of eigenvectors.

After enough simplifications for $U(2)$ coin with eigenvectors \eqref{e_k}, we will have

\begin{equation}\label{C(k)}
\mathcal{C}(k)_{U(2)}=\left( \begin {array}{cccc} 1-L & -F^* & -F^* & G^* \\\noalign{\medskip}-F&L &L &F^* \\\noalign{\medskip}-F&L &L & F^* 
\\\noalign{\medskip}G&F&F& 1-L \end {array}
\right) 
\end{equation}
with
\begin{align}
L&=\frac{\sin^2\theta}{2\sin^2\gamma}\\\nonumber
G&=-\frac{\sin^2\theta}{2\sin^2\gamma}{\rm e}^{2i \left( k-\beta \right)}\\\nonumber
F&=\frac{i\sin(k-\alpha)\sin\theta\cos\theta }{2\sin^2\gamma}{\rm e}^{i \left( k-\beta \right)}
\end{align}
where
\begin{equation}
\cos{\gamma}=\cos{\theta}\cos{\left(k-\alpha\right)}.
\end{equation}

$\mathcal{C}(k)_{U(2)}$, determines the asymptotic reduced density matrix of 1DQW with general coin of $U(2)$. Specially for local initial states, we can use \eqref{C_L} to find
\begin{equation}\label{C_LU(2)}
\mathcal{C}^{L}_{U(2)}=\frac{1}{2}\left( \begin {array}{cccc} 2-\sin{\theta} & f^* & f^* & g^* \\\noalign{\medskip}f&\sin{\theta} &\sin{\theta} &-f^* \\\noalign{\medskip}f&\sin{\theta} &\sin{\theta} & -f^* 
\\\noalign{\medskip}g&-f&-f&2-\sin{\theta} \end {array}
 \right) 
\end{equation}
where
\begin{align}\label{f-g-local}
f&=\frac{\sin{\theta} \cos{\theta}}{\sin{\theta}+1}{\rm e}^{i \left( \alpha-\beta \right) } \\\nonumber
g&=\frac {\sin{\theta}\left( \sin{\theta}-1\right) }{\sin{\theta}+1}\,{\rm e}^{2\,i \left( \alpha-\beta \right) }
\end{align}

As it seems the phase $\alpha$ and $\beta$ appear only in the form of $\alpha-\beta$ in $\mathcal{C}^L_{U(2)}$, So the long time behavior of reduced density matrix of local initial states only depends on the difference of phases not the individual value of them.
\section{Examples}
In this section we use the formalism of previous section in order to investigate some asymptotic behaviors of QW with different kinds of initial states.
\subsection{Local initial state}
Assume an initial state localized at the origin ($x=0$) with coin state $\ket{\chi_0}=\ket{0}$. So from \eqref{P0} we have
\begin{equation}\label{P_0_L}
P_k\left(0\right)=\opp{0}{0}=\left( \begin {array}{cc} 1&0\\\noalign{\medskip}0&0\end {array} \right).
\end{equation}
By plug it in \eqref{rho_c-local} and using explicit form of $\mathcal{C}^{L}_{U(2)}$ in \eqref{C_LU(2)}, easily we can find asymptotic reduced density matrix $\hat\rho_c$ as
\begin{equation}\label{rho_c-local-init0}
\hat\rho_c=\frac{1}{2}\left( \begin {array}{cc} 2-\sin \left( \theta \right) &{f^*}
\\ {f}&\sin \left( \theta \right) 
\end {array} \right),
\end{equation}
where $f$ has been defined in \eqref{f-g-local}. This $\hat\rho_c$ can be used to investigate some features of QW such as coin-position entanglement (CPE) \cite{annabestani}, entanglement temperature \cite{Romanelli2012} or chairality probability distribution \cite{Romanelli2014}. As an example the CPE can be calculated by Von-neumann entropy 
\begin{equation}\label{E_0}
E=-\sum_{i=1}^2{\lambda_i\log_2{\lambda_i}},
\end{equation}
where $\lambda_i$s are eigenvalues of $\hat{\rho_c}$. From \eqref{rho_c-local-init0} it is easy to see that 
\begin{equation}
\lambda_1,\lambda_2=\frac{1}{2}\pm\frac{\left|\cos{\theta}\right|}{2\sin{\theta}+2}
\end{equation}

\begin{figure}
\includegraphics[width=7cm]{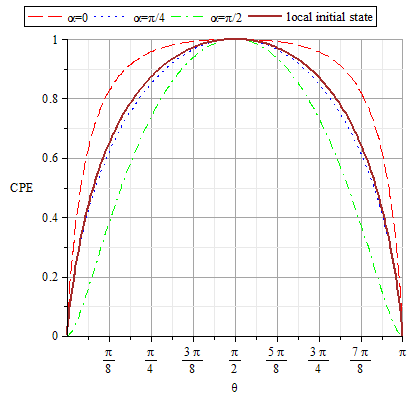}
\caption{CPE for 1DQW with $U(2)$ coin operator for locla initial state \eqref{P_0_L} (solid thick line). CPE for non-local separable initial state \eqref{nonlocal-separable} for $\alpha=0$ (dash red line), $\alpha=\pi/4$ (dot blue line) and $\alpha=\pi/2$ (dash-dot green line).}
\label{Fig:CPE_compare}
\end{figure}

The Fig.\ref{Fig:CPE_compare} shows the CPE for initial state $\ket{0}$. Note that CPE for $\theta=\pi/4$ is 0.872 as we expected and has been calculated before in Ref.\cite{Carneiro,Abal06-ent} for Hadamard walk (Note that $U_{\pi/4,\pi/2,\pi/2}=H $ is Hadamard coin). Furthermore we can see $\lambda_i $ does not depend on phase $\alpha$ and $\beta$. One may ask "Is it universal feature for local initial state?". Answer to this question is easy in our formalism. Let us assume most general form of initial state as 
\begin{equation}
\ket{\psi}_g=\cos{\left(\frac{\xi}{2}\right)}\ket{0}+e^{i\eta}\sin{\left(\frac{\xi}{2}\right)}\ket{1}
\end{equation} 
which is localized at $x=0$. After some calculation and simplification, the form of eigenvalues will be
\begin{equation}
\frac{1}{2}\pm{\frac {\sqrt {1+\cos \left( 2\theta \right) \cos \left( 2\xi \right) +\sin \left( 2\theta \right) \cos \left(\alpha-\beta-\eta \right) \sin \left( 2\xi \right) }}{2\sqrt {2}\left(\sin \left( 
		\theta \right) +1\right)}}.
\end{equation}
It means that for the initial states $\ket{0}, \ket{1}$ ($\xi=0$, $\xi=\pi$ respectively) and equally superposed initial state ($\xi=\frac{\pi}{2}$), even for most general form of coin operator, we don't have dependency on phases $\alpha$ and $\beta$. On the other side for any initial states, if the coin operator be one of the Pauli matrices $\sigma_i$ with $i=0\dots 3$ ($\theta=0,\pi,\frac{\pi}{2}$), phases $\alpha$ and $\beta$ do not appear in the eigenvalues. 
\subsection{Separable distributed initial state}
\begin{figure}
	\includegraphics[width=10cm]{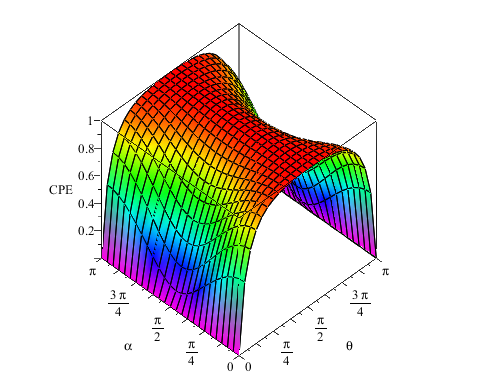}
	\caption{Entanglement between coin and position for 1DQW with $U(2)$ coin operator for the initial state \eqref{nonlocal-separable}}
	\label{Fig:CPE_3D}
\end{figure}

Assume an initial state $\ket{\psi}_d$ with coin state $\ket{\chi_0}=\ket{0}$ which is distributed over $x$ axis at $x=\pm 1$ with equal weight
\begin{equation}\label{nonlocal-separable}
\ket{\psi}_d=\frac{1}{\sqrt{2}}\left(\ket{-1}_x+\ket{1}_x\right)\otimes \ket{0}.
\end{equation}
So from \eqref{Qk} we have
\begin{equation}
Q(k)=\frac{1}{\sqrt{2}}\left(e^{-ik}+e^{ik}\right)=\sqrt{2}\cos{k}.
\end{equation}
By putting $Q(k)$ in \eqref{C_S}, taking integral and using \eqref{rho_c-separable}, we will have
\begin{equation}\label{rho_c-distributed-init0}
\hat\rho_c=\frac{\sin\left(\theta\right)}{\sin \left( \theta \right) +1}\left( \begin {array}{cc} \frac{\sin \left( \theta \right) +1}{\sin\left(\theta\right)}-B &{A}
\\ {A^*}&B 
\end {array} \right),
\end{equation}
where
\begin{align}
A&=\cos \left( \theta \right) \left( \sin \left(\alpha \right)^{2}+{\frac {1/2-i\sin \left( \alpha \right){{\rm e}^{-i\alpha}}}{\sin \left( \theta \right) +1}} \right){\rm e}^{i\left(\beta-\alpha \right)} \\\nonumber
B&={\sin\left( \theta \right)\sin \left( \alpha \right)^{2}+ \cos \left(\alpha \right)^{2} }.
\end{align}
The eigenvalues of $\hat\rho_c$ are
\begin{equation}
\frac{1}{2}\pm{\frac { \left| \cos \left( \theta \right)  \right| \sqrt {4\sin \left( \theta \right)^{2}\sin \left(\alpha \right)^{2}+4\sin\left( \theta \right)\sin \left( \alpha \right)^{2}+1}}{ 2\left( \sin \left(\theta \right) +1 \right) ^{2}}}.
\end{equation}
so by using \eqref{E_0}, the CPE is completely determined. CPE has been plotted in Fig.\ref{Fig:CPE_3D} and Fig.\ref{Fig:CPE_compare}
\subsection{Entangled initial state}
As a final example, assume that initial state be an entangled state as
\begin{equation}\label{entangled_initalStae}
\ket{\psi}_d=\frac{1}{\sqrt{2}}\left(\ket{-1}_x\otimes \ket{0}+\ket{1}_x\otimes \ket{1}\right).
\end{equation}
Using \eqref{FourierTransform} and \eqref{psi_k}, we have
\begin{align}
P_k\left(0\right)=\frac{1}{2}\left( \begin {array}{cc} 1 & {\rm e}^{2ik} 
\\\noalign{\medskip} {{\rm e}^{-2ik}} &  1 \end {array} \right)
\end{align}
\begin{figure}
\includegraphics[width=7cm]{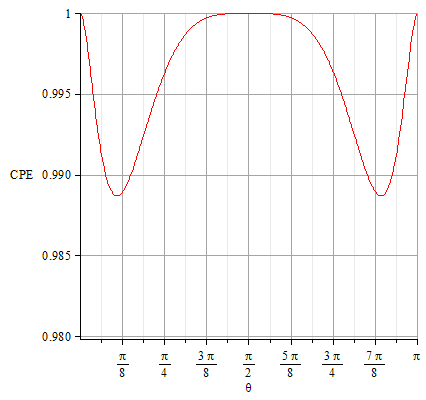}
\caption{CPE for 1DQW with $U(2)$ coin operator for the entangled initial state \eqref{entangled_initalStae}. Note that, the vertical axis restricted to [0.98,1] for a better view }\label{Fig:CPE_entangled}
\end{figure}
then finding $\tilde{\rho_c}$ from \eqref{rho_c-trace-Form} and it's eigenvalues will be simple.
 \begin{align}
\lambda_1,\lambda_2= \frac{1}{2}\pm{\frac {\sin \left( \theta \right)  \left( 1-\sin \left( 
 		\theta \right)\right) }{2\left( 1+\sin \left( \theta \right) 
 		\right) ^{2}}}
 \end{align}
 
 CPE have been plotted in Fig.\ref{Fig:CPE_entangled}. We can see for the entangled initial state (\ref{entangled_initalStae}), CPE only depends on $\theta$ and for almost all values of $\theta$ is maximum. This CPE decays just in two points near to  $\pi/8$ and $7\pi/8$ to 0.99 (see Fig.\ref{Fig:CPE_entangled}). 
 \section{conclusion}
 In this paper we have shown that, asymptotic reduced density matrix of any quantum walk with \textit{m}-dimensional coin space, can be completely determined by $m^2\times m^2$ specific matrix $\mathcal{C}(\mathbf{k})$ with $\hat\rho_c=\int\frac{d^m\mathbf{k}}{(2\pi)^m} {Tr_1\left(P_0(\mathbf{k})\otimes I\;\mathcal{C}(\mathbf{k})\right)}$.
 The characteristic matrix $\mathcal{C}(\mathbf{k})$ depends only on the eigenvalues of the evolution operator, with special symmetry (see (\ref{C})). Our calculations show that this symmetry helps us to simplify final form of $\mathcal{C}(\mathbf{k})$ to have clean compact expressions for almost all types of quantum walks.
 Although finding the clean form of $\mathcal{C}(\mathbf{k})$ requires relatively tedious calculations and simplifications, it is always possible and only needed once.
 
 We have calculated $\mathcal{C}(\mathbf{k})$ for most general form of coin operator in one dimensional quantum walk in this paper as well and tried to show the power of our formalism with several examples. Although some of the results of our examples are new and have not been addressed before, they are not the main results of our paper, the main aim of this paper is introducing characteristic matrix approach to calculate asymptotic reduced density matrix. This approach have been introduced by 
 (\ref{rho_c-trace-Form})   and (\ref{C}). One can use (\ref{C}) to calculate $\mathcal{C}(\mathbf{k})$ for a specific type of QW, thereafter for any initial state, asymptotic reduced density matrix of QW is completely known by (\ref{rho_c-trace-Form}). So this formalism can be used as a powerful tool for analytic study of asymptotic behavior of QW, related to reduced density matrix.
 
  One of the important quantities, related to reduced density matrix, is coin-position entanglement which has been studied here for 1DQW with general U(2)  coin operator just as examples. But other quantities such as entanglement temperature, chirality probability distribution, etc can be easily studied by this formalism.
  
  We hope this method  can help researchers to have better and easier study of asymptotic behavior of quantum walks.

\bibliography{qw}

\end{document}